
%
%
\input harvmac %
%
%
%
%
%
%
%
%
%
\newif\ifdraft

\noblackbox
\catcode`\@=11
\newif\iffrontpage
%
\ifx\answ\bigans
\def\titleft{\titsm}
\magnification=1200\baselineskip=15pt plus 2pt minus 1pt
%
\advance\hoffset by-0.075truein
\hsize=6.15truein\vsize=600.truept\hsbody=\hsize\hstitle=\hsize
\else\let\lr=L
\def\titleft{\titla}
\magnification=1000\baselineskip=14pt plus 2pt minus 1pt
%
\vsize=6.5truein
\hstitle=8truein\hsbody=4.75truein
\fullhsize=10truein\hsize=\hsbody
\fi
\parskip=4pt plus 10pt minus 4pt

\font\titla=cmr10 scaled\magstep3
\font\tenmss=cmss10
\font\absmss=cmss10 scaled\magstep1
\newfam\mssfam
\font\footrm=cmr8  \font\footrms=cmr5
\font\footrmss=cmr5   \font\footi=cmmi8
\font\footis=cmmi5   \font\footiss=cmmi5
\font\footsy=cmsy8   \font\footsys=cmsy5
\font\footsyss=cmsy5   \font\footbf=cmbx8
\font\footmss=cmss8
\def\footfont{\def\rm{\fam0\footrm}
\textfont0=\footrm \scriptfont0=\footrms
\scriptscriptfont0=\footrmss
\textfont1=\footi \scriptfont1=\footis
\scriptscriptfont1=\footiss
\textfont2=\footsy \scriptfont2=\footsys
\scriptscriptfont2=\footsyss
\textfont\itfam=\footi \def\it{\fam\itfam\footi}
\textfont\mssfam=\footmss \def\mss{\fam\mssfam\footmss}
\textfont\bffam=\footbf \def\bf{\fam\bffam\footbf} \rm}
\def\tenpoint{\def\rm{\fam0\tenrm}
\textfont0=\tenrm \scriptfont0=\sevenrm
\scriptscriptfont0=\fiverm
\textfont1=\teni  \scriptfont1=\seveni
\scriptscriptfont1=\fivei
\textfont2=\tensy \scriptfont2=\sevensy
\scriptscriptfont2=\fivesy
\textfont\itfam=\tenit \def\it{\fam\itfam\tenit}
\textfont\mssfam=\tenmss \def\mss{\fam\mssfam\tenmss}
\textfont\bffam=\tenbf \def\bf{\fam\bffam\tenbf} \rm}
\ifx\answ\bigans\def\abstractfont{\tenpoint}\else
\def\abstractfont{\def\rm{\fam0\absrm}
\textfont0=\absrm \scriptfont0=\absrms
\scriptscriptfont0=\absrmss
\textfont1=\absi \scriptfont1=\absis
\scriptscriptfont1=\absiss
\textfont2=\abssy \scriptfont2=\abssys
\scriptscriptfont2=\abssyss
\textfont\itfam=\bigit \def\it{\fam\itfam\bigit}
\textfont\mssfam=\absmss \def\mss{\fam\mssfam\absmss}
\textfont\bffam=\absbf \def\bf{\fam\bffam\absbf}\rm}\fi
%
\def\f@@t{\baselineskip10pt\lineskip0pt\lineskiplimit0pt
\bgroup\aftergroup\@foot\let\next}
\setbox\strutbox=\hbox{\vrule height 8.pt depth 3.5pt width\z@}
\def\vfootnote#1{\insert\footins\bgroup
\baselineskip10pt\footfont
\interlinepenalty=\interfootnotelinepenalty
\floatingpenalty=20000
\splittopskip=\ht\strutbox \boxmaxdepth=\dp\strutbox
\leftskip=24pt \rightskip=\z@skip
\parindent=12pt \parfillskip=0pt plus 1fil
\spaceskip=\z@skip \xspaceskip=\z@skip
\Textindent{$#1$}\footstrut\futurelet\next\fo@t}
\def\Textindent#1{\noindent\llap{#1\enspace}\ignorespaces}
\def\footnote#1{\attach{#1}\vfootnote{#1}}%

\def\foot{\attach\footsymbolgen\vfootnote{\footsymbol}}
\let\footsymbol=\star
\newcount\lastf@@t           \lastf@@t=-1
\newcount\footsymbolcount    \footsymbolcount=0
\def\footsymbolgen{\relax\footsym
\global\lastf@@t=\pageno\footsymbol}
\def\footsym{\ifnum\footsymbolcount<0
\global\footsymbolcount=0\fi
{\iffrontpage \else \advance\lastf@@t by 1 \fi
\ifnum\lastf@@t<\pageno \global\footsymbolcount=0
\else \global\advance\footsymbolcount by 1 \fi }
\ifcase\footsymbolcount \fd@f\star\or
\fd@f\dagger\or \fd@f\ast\or
\fd@f\ddagger\or \fd@f\natural\or
\fd@f\diamond\or \fd@f\bullet\or
\fd@f\nabla\else \fd@f\dagger
\global\footsymbolcount=0 \fi }
\def\fd@f#1{\xdef\footsymbol{#1}}
\def\space@ver#1{\let\@sf=\empty \ifmmode #1\else \ifhmode
\edef\@sf{\spacefactor=\the\spacefactor}
\unskip${}#1$\relax\fi\fi}
\def\attach#1{\space@ver{\strut^{\mkern 2mu #1}}\@sf}
%
\newif\ifnref
\def\rrr#1#2{\relax\ifnref\nref#1{#2}\else\ref#1{#2}\fi}
\def\ldf#1#2{\begingroup\obeylines
\gdef#1{\rrr{#1}{#2}}\endgroup\unskip}
\def\nrf#1{\nreftrue{#1}\nreffalse}
\def\doubref#1#2{\refs{{#1},\ {#2}}}

\nreffalse
\def\refout{\listrefs}
%
\def\eqn#1{\xdef #1{(\secsym\the\meqno)}
\writedef{#1\leftbracket#1}%
\global\advance\meqno by1\eqno#1\eqlabeL#1}
\def\eqnalign#1{\xdef #1{(\secsym\the\meqno)}
\writedef{#1\leftbracket#1}%
\global\advance\meqno by1#1\eqlabeL{#1}}
%
\def\chap#1{\newsec{#1}}
\def\chapter#1{\chap{#1}}
\def\sect#1{\subsec{{ #1}}}
\def\section#1{\sect{#1}}
\def\\{\ifnum\lastpenalty=-10000\relax
\else\hfil\penalty-10000\fi\ignorespaces}
\def\note#1{\leavevmode%
\edef\@@marginsf{\spacefactor=\the\spacefactor\relax}%
\ifdraft\strut\vadjust{%
\hbox to0pt{\hskip\hsize%
\ifx\answ\bigans\hskip.1in\else\hskip-.1in\fi%
\vbox to0pt{\vskip-\dp
\strutbox\sevenbf\baselineskip=8pt plus 1pt minus 1pt%
\ifx\answ\bigans\hsize=.7in\else\hsize=.35in\fi%
\tolerance=5000 \hbadness=5000%
\leftskip=0pt \rightskip=0pt \everypar={}%
\raggedright\parskip=0pt \parindent=0pt%
\vskip-\ht\strutbox\noindent\strut#1\par%
\vss}\hss}}\fi\@@marginsf\kern-.01cm}
\def\titlepage{%
\frontpagetrue\nopagenumbers\abstractfont%
\hsize=\hstitle\rightline{\vbox{\baselineskip=10pt%
{\abstractfont\pubnum}}}\pageno=0}
\frontpagefalse
\def\pubnum{}
\def\pdate{\number\month/\number\yearltd}
\def\makefootline{\iffrontpage\vskip .27truein
\line{\the\footline}
\vskip -.1truein\leftline{\vbox{\baselineskip=10pt%
{\abstractfont\pdate}}}
\else\vskip.5cm\line{\hss \tenrm $-$ \folio\ $-$ \hss}\fi}
\def\title#1{\vskip .7truecm\titlestyle{\titleft #1}}
\def\titlestyle#1{\par\begingroup \interlinepenalty=9999
\leftskip=0.02\hsize plus 0.23\hsize minus 0.02\hsize
\rightskip=\leftskip \parfillskip=0pt
\hyphenpenalty=9000 \exhyphenpenalty=9000
\tolerance=9999 \pretolerance=9000
\spaceskip=0.333em \xspaceskip=0.5em
\noindent #1\par\endgroup }
\def\autskip{\ifx\answ\bigans\vskip.5truecm\else\vskip.1cm\fi}
\def\author#1{\vskip .7in \centerline{#1}}

\def\address#1{\ifx\answ\bigans\vskip.2truecm
\else\vskip.1cm\fi{\it \centerline{#1}}}
\def\abstract#1{
\vskip .3in\vfil\centerline
{\bf Abstract}\penalty1000
{{\smallskip\ifx\answ\bigans\leftskip 2pc \rightskip 2pc
\else\leftskip 5pc \rightskip 5pc\fi
\noindent\abstractfont \baselineskip=12pt
{#1} \smallskip}}
\penalty-1000}
\def\endpage{\tenpoint\supereject\global\hsize=\hsbody%
\frontpagefalse\footline={\hss\tenrm\folio\hss}}
\def\ack{\goodbreak\vskip2.cm\noindent {\bf Acknowledgements}}

\def\bfone{\relax{\rm 1\kern-.35em 1}}
\def\inbar{\vrule height1.5ex width.4pt depth0pt}
\def\IC{\relax\,\hbox{$\inbar\kern-.3em{\mss C}$}}
\def\ID{\relax{\rm I\kern-.18em D}}
\def\IF{\relax{\rm I\kern-.18em F}}
\def\IH{\relax{\rm I\kern-.18em H}}
\def\II{\relax{\rm I\kern-.17em I}}
\def\IN{\relax{\rm I\kern-.18em N}}
\def\IP{\relax{\rm I\kern-.18em P}}
\def\IQ{\relax\,\hbox{$\inbar\kern-.3em{\rm Q}$}}
\def\IR{\relax{\rm I\kern-.18em R}}
\font\cmss=cmss10 \font\cmsss=cmss10 at 7pt
\def\ZZ{\relax\ifmmode\mathchoice
{\hbox{\cmss Z\kern-.4em Z}}{\hbox{\cmss Z\kern-.4em Z}}
{\lower.9pt\hbox{\cmsss Z\kern-.4em Z}}
{\lower1.2pt\hbox{\cmsss Z\kern-.4em Z}}\else{\cmss Z\kern-.4em
Z}\fi}
  \def\d{\delta}
\def\e{\epsilon} 
 \def\l{\lambda}
 \def\s{\sigma}

\def\cL{{\cal L}}

\def\nup#1({Nucl.\ Phys.\ $\us {B#1}$\ (}
\def\plt#1({Phys.\ Lett.\ $\us  {#1}$\ (}
\def\cmp#1({Comm.\ Math.\ Phys.\ $\us  {#1}$\ (}
\def\prp#1({Phys.\ Rep.\ $\us  {#1}$\ (}
\def\prl#1({Phys.\ Rev.\ Lett.\ $\us  {#1}$\ (}
\def\prv#1({Phys.\ Rev.\ $\us  {#1}$\ (}
\def\mpl#1({Mod.\ Phys.\ Let.\ $\us  {#1}$\ (}
\def\tit#1|{{\it #1},\ }
%

%

\def\tilde{\widetilde}
\def\bar{\overline}
\def\us#1{\underline{#1}}

\def\Coe#1.#2.{{#1\over #2}}
\def\coeff#1#2{\relax{\textstyle {#1 \over #2}}\displaystyle}
\def\coe#1.#2.{\relax{\textstyle {#1 \over #2}}\displaystyle}

\def\to{\rightarrow}
\def\notin{\hbox{{$\in$}\kern-.51em\hbox{/}}}

\def\del{\partial}

 \def\ie{{i.e.}}
\catcode`\@=12
%
%
\def\brk{\hfill\break}
\def\Mgut{M_{\rm GUT}}

\def\Mweak{M_{Z}}
\def\Lsoft{\cL_{\rm soft}}

\def\mg{\tilde m}
\def\mgzero{\mg_{1/2}}
\def\mgrav{ m_{3/2}}
\def\mav{M_{\rm av}}
\def\mavzero{m_{\rm av}}
\def\Mm{M}
\def\DMm{\Delta \Mm}
\def\xzero{x_0}

\def\hu{h_{u}}
\def\hd{h_{d}}
\def\q{\tilde q_L}
\def\qb{\bar{\tilde q}_L}
\def\u{\tilde u_R}
\def\ub{\bar{\tilde u}_R}
\def\d{\tilde d_R}
\def\db{\bar{\tilde d}_R}
\def\l{\tilde l_L}
\def\lb{\bar{\tilde l}_L}
\def\e{\tilde e_R}
\def\eb{\bar{\tilde e}_R}

\def\ytop{Y_{\rm t}}

\def\Au{A^{u}}
\def\Ad{A^{d}}
\def\Al{A^{l}}
\def\hc{h.c.}
\def\Mz{M_Z}
\def\s2w{\sin^2 \theta_w}

\def\dt{\del_t}

\def\gsim{\mathrel{\mathpalette\Vversim>}}
\def\Vversim#1#2{\vcenter{\offinterlineskip
        \ialign{$#1\hfil##\hfil$\crcr#2\crcr\sim\crcr } }}
%
%
%
\ldf\who{Who should be referenced here????????????}
\ldf\fcncearly{
J.~Ellis and D.V.~Nanopoulos, \plt 110B (1982) 44; \brk
R.~Barbieri and R.~Gatto, \plt 110B (1982) 211;\brk
M.~Duncan, \nup 221 (1983) 285;\brk
J.~Donoghue, H.-P.~Nilles, and D.~Wyler, \plt B128 (1983) 55;\brk
A.Bouquet, J.~Kaplan and C.A.~Savoy, \plt B148 (1984) 69.}
\ldf\HKR{L.~Hall, V.~Kostelecky and S.~Raby, \nup 267 (1986) 415.}
\ldf\BBMR{S.~Bertolini, F.~Borzumati, A.~Masiero and G.~Ridolfi,
\nup 353 (1991) 591.}
\ldf\GM{F.~Gabbiani and  A.~Masiero, \nup 322 (1989) 235.}
\ldf\NS{Y.~Nir and N.~Seiberg, \plt B309 (1993) 337.}
\ldf\DLK{M.~Dine, R.~Leigh and A.~Kagan, \prv D84 (1993) 4269.}
%
\ldf\DKS{M.~Dine, A.~Kagan and S.~Samuel, \plt 243B (1990) 250.}
%
\ldf\BIM{A.~Brignole, L.~Ib\'a\~nez and C.~Mu\~noz, preprint FTUAM-26/93.}
\ldf\BLM{R.~Barbieri, J.~Louis and M.~Moretti,
\plt B312 (1993) 451.}
\ldf\IL{L.~Ib\'a\~nez and D.~L\"ust, \nup382 (1992) 305.}
\ldf\KLa{V.~Kaplunovsky and J.~Louis, \plt B306 (1993) 269.}
\ldf\ILM{L.~Ib\'a\~nez and  C.~Lopez, \nup 233 (1984) 511; \brk
L.~Ib\'a\~nez, C.~Lopez and C.~Mu\~noz, \nup 256 (1985) 218.}
\ldf\SoWe{Soni and Weldon}
\ldf\BFSW{
R.~Barbieri, S.~Ferrara and C.~Savoy, \plt 119B (1982) 343;\brk
S.~K.~Soni and H.~A.~Weldon, \plt 126B (1983) 215.}
\ldf\HKT{J.~Hagelin, S.~Kelley, T.~Tanaka, \nup 415 (1994) 293;
\mpl{A8}   (1993) 2737.}
\ldf\nillesrep{For a review see for example,
H.-P.~Nilles, \prp  C110 (1984) 1 and references therein.}
\ldf\IL{L.~Ib\'a\~nez and D.~L\"ust, \nup382 (1992) 305.}
\ldf\BMMP{A.~Buras, M.~Misiak, M.~Munz and S.~Pokorski, Max-Planck preprint
MPI-PH-93-77.}
\ldf\PD{K.~Hikasa et al., Particle Data Group \prv{D45} (1992) 1.}
\ldf\LN{J.~Louis and Y.~Nir, in preparation.}
\ldf\MN{
A.~Lleyda and C.~Mu\~noz, \plt 317 (1993) 82;\brk
N.~Polonsky and A.~Pomarol,
University of Pennsylvania preprint UPR-0616-T (1994);\brk
D.~Matalliotakis and H.P.~Nilles, Munich preprint TUM-HEP-201/94;\brk
M.~Olechowski and S.~Pokorski, Munich preprint MPI-PHT/94-40.}
\ldf\Frank{F.~Eberlein, Diplomthesis, Universit\"at M\"unchen.}
\ldf\Anja{A.~K\"onig, Diplomthesis, Universit\"at M\"unchen.}
\ldf\Banks{T.~Banks, \nup 303 (1988) 172.}
\ldf\cleo{A.~Bean et al., \prl 70 (1993) 138.}
\ldf\twenty{K.~Inoue, A.~Kakuto, H.~Komatsu and S.~Takeshita,
Prog.~Theor.~Phys. $\us{67}$ (1982) 889; $\us{68}$ (1983) 927;\brk
J.P.~Derendinger and C.A.~Savoy, \nup 237 (1984) 307;\brk
N.K.~Falck, Z.~Phys. C30 (1986) 247.}
\ldf\COWE{L.~Ib\'a\~nez   and  G.~Ross \plt 110 (1982) 215; \brk
K.~Inoue  A.~Kakuto, H.~Komatsu and S.~Takeshita,
 Prog.~Theor.~Phys.  $\us{68}$ (1983) 927;\brk
L.~Alvarez-Gaum\'e, J.~Polchinsky and M.~Wise, \nup 221 (1983) 495.}
\ldf\DGH{M.~Dugan, B.~Grinstein and L.~Hall, \nup 255 (1985) 413.}
%
%
\def\aff#1#2{\centerline{$^{#1}${\it #2}}}
\def\pubnum{\hbox{MPI--PhT/94-51}\hbox{LMU--TPW 94-12}
\hbox{July 1994}}
\titlepage
\vskip 2cm
\title{Constraints on non-universal soft terms
from flavor changing neutral currents}
\author{D.~Choudhury$^{1}$, F.~Eberlein$^{2}$, A.~K\"onig$^{2}$,
J.~Louis$^{2}$\foot{Supported by  a Heisenberg fellowship of the DFG.}
 and S.~Pokorski$^{1,3}$}
\vskip2.truecm
\aff1{Max-Planck Institut f\"ur Physik,
 F\"ohringer Ring 6, D-80805 M\"unchen, Germany }
\aff2{Sektion Physik, Universit\"at M\"unchen, \break
Theresienstr.~37, D-80333 M\"unchen, Germany}
\aff3{Institute of Theoretical Physics, Warsaw University, \break
 ul. Hoza 69,
00-681 Warsaw, Poland}
\vskip 1.5cm
\abstract{
The smallness of flavor changing neutral currents  constrains
the soft parameter space of supersymmetric extensions of the Standard Model.
These low energy constraints are
translated to the soft parameter space generated
at some high energy scale $\Mgut$.
For gaugino masses larger than the scalar masses
and  non-universal $A$-terms  the constraints
are significantly diluted at $\Mgut$ and do
 allow for the possibility
of non-universal scalar masses.
The strongest constraints arise in the slepton sector of the theory.
  }
\endpage
%
%
%
%
{\bf 1}.
The successful prediction of the smallness of flavor changing
neutral currents (FCNC)  is one of the
cornerstones of the Standard Model (SM).
Most extensions of the SM contain new sources for FCNC
  which lead to  a
delicate test of any new physics  above  the weak scale $\Mweak$.
 $N=1$ supersymmetric versions of the SM appear to be   promising
candidates for such new physics and most supersymmetric models do contain
additional  contributions to FCNC via gaugino
 exchange in box and/or penguin
diagrams \doubref\fcncearly\HKR.
The measurements of neutral meson mixing
and radiative decays \doubref\PD\cleo
$$
\eqalign{
{\Delta m_K \over m_K}\ &=\ 7 \times 10^{-15} \ , \qquad
{\Delta m_B \over m_B}\ =\ 7 \times 10^{-14} \ , \qquad
{\Delta m_D \over m_D}\ \le\ 7 \times 10^{-14} \ , \cr
&BR(B\to X_s \gamma)  \le \  5.4 \times 10^{-4} \ , \quad
BR(\mu\to e\ \gamma)  \le \  5 \times 10^{-11} \ , \cr
&BR(\tau\to \mu\ \gamma)  \le \  4.2 \times 10^{-6} \ , \quad
BR(\tau \to e\ \gamma)  \le \  2 \times 10^{-4} \ ,
}
\eqn\bounds
$$
impose severe constraints on the (off-diagonal elements
of the) mass matrix of the
squarks and sleptons \refs{\fcncearly, \GM, \HKT}.

The masses of the squarks and sleptons arise as
a consequence of supersymmetry and electroweak
symmetry breaking.
At present  soft  breaking of supersymmetry
seems to be
the most attractive  mechanism for
 generating
 a  phenomenologically acceptable  scalar mass spectrum.
Such softly broken supersymmetric theories
appear rather naturally
in the flat limit of spontaneously broken
  supergravity models where
 a hidden sector induces the breakdown and
gravitational interactions communicate
the breaking to the observable sector.
 This mechanism  induces
 soft  terms in  the observable sector
(which contains the quark and lepton supermultiplets)
 at some high energy scale $\Mgut$.
However, the constraints \bounds\
hold at the weak scale $\Mweak$ and
cannot be  directly applied to the soft parameter
space at high energies. Instead,
renormalization corrections of the scalar masses
have to be taken into account. As a consequence,
 the scalar masses
at  low energies
can be (significantly) different  from the soft input parameters
generated at
high  energies ($\Mgut$) if large quantum corrections are present.
The prime example of this phenomenon is the supersymmetric
Coleman-Weinberg mechanism where renormalization effects turn
a Higgs mass parameter negative and radiatively  induce
 the electroweak symmetry breakdown \COWE.

In  the simplest supergravity models  \nillesrep\ the soft terms
are universal at $\Mgut$:
all scalar masses $m^2_{i j} $
are determined by  the gravitino mass
$\mgrav$ ($m_{i j}^2 = \delta_{i j}\ \mgrav^2$)
and the trilinear scalar couplings $A_{ij} $
are proportional to the corresponding Yukawa couplings $Y_{ij}$
($A_{ij} = A\ \mgrav Y_{ij}$).
Renormalization corrections to the scalar masses
indeed induce some small non-universality at $\Mweak$
but nevertheless the current bounds \bounds\ are satisfied
\refs{\BBMR, \GM, \HKT}.
In more generic models
(particularly in many  string models
\nrf{\IL\KLa\BIM}\refs{\IL,\KLa,\BIM})
 non-universal soft terms do arise at
$\Mgut$\nrf{\BFSW} \refs{\BFSW, \HKR, \GM} and
from the point of view of (string-) model building
it is of  interest to analyze
to what extent such non-universality
 can be tolerated without violating the low energy
data \bounds.
In other words one would like
to translate the bounds \bounds\ into constraints
on the soft parameter space generated at $\Mgut$.
We find  two possible sources which can significantly dilute
the bounds.
On the one hand, large gaugino masses enhance the diagonal
or `average' squark masses $\mav$\nrf{\DKS} \refs{\DKS,\BIM}
whereas  non-universal $A$-terms
can  decrease the off-diagonal mass matrix elements.
Both effects together lead to a dilution of the constraints at $\Mgut$
and allow for  non-universal soft terms
 at the high energy scale.\foot{
Other aspects of non-universal scalar mass terms have recently
been discussed in refs.~\refs{\NS, \DLK, \MN}.}
It is the purpose of this letter to make these statements
 more quantitative.\foot{
An extended version of this letter can be found in
refs.~\doubref\Frank\Anja.}

%
%

%
{\bf 2}.
In supersymmetric extensions of the SM
all chiral fermions are promoted to chiral
$N=1$ supermultiplets and one additional Higgs doublet is
 introduced.
Let us assume that
supersymmetry is  broken by
  generic soft terms at some high energy scale $\Mgut$
$$
\eqalign{
- \Lsoft\ &=\  \mgrav\
\big( \Au_{ij}\  \u^j \q^i\hu + \Ad_{ij}\  \d^j \q^i\hd
+ \Al_{ij}\  \e^j \l^i\hd
+ \hc \big) \cr
+\ &(m^2_q)_{ij}\, \q^i \qb^j + (m^2_u)_{ij}\, \u^i \ub^j
+ (m^2_d)_{ij}\, \d^i \db^j
+ (m^2_l)_{ij}\, \l^i \lb^j + (m^2_e)_{ij}\, \e^i \eb^j\  \cr
 +\ & m_1^2\, |\hu|^2 +  m_2^2\, |\hd|^2  + ( B\ \mgrav^2\, \hu \hd +
\sum_{a=1}^3 \coeff12 \mg_a  (\lambda\lambda)_a + \hc)  \ ,}
\eqn\defLsoft
$$
where $i,j$  are  summed over $1,2,3$. $\q$ ($\u,\d$) denotes
 the left- (right-) handed  squarks,
 $\l$ ($\e$)   the left- (right-) handed sleptons and
$\hu, \hd$  the two Higgs doublets;
$\mg_a$
are  the three gaugino masses
of $SU(3), SU(2), U(1)$ respectively
and for simplicity we take them to be universal (at $\Mgut$)
 $\mg_1 = \mg_2 = \mg_3 = \mgzero$.\foot{
In general, the soft terms \defLsoft\  introduce  new (and dangerous)
CP-violating phases  \DGH. We do not address the related
phenomenological problems
 but instead assume
that all possible sources of CP-violation  are small.
(The constraints on the soft parameter space
 usually become  stronger
when arbitrary phases of $O(1)$  are present.)}

The soft terms displayed in eq.~\defLsoft\ determine the
scalar masses at the weak scale after their renormalization
corrections
 (between $\Mgut$ and $\Mweak$)
have  been
 taken into account.
   These corrections are determined by the  solutions of the
appropriate  (one-loop) renormalization group (RG)
equations  \refs{\twenty, \ILM, \BBMR}.
For the purpose of this paper, it is
instructive to first work in an approximation where
 only  the  gauge couplings and  $A$-terms are kept
while  all Yukawa couplings are neglected in the RG-equations.
This simplifies the discussion considerably
and clearly shows  the physical effects involved.
The `zero Yukawa limit'
 is a very good approximation except for the top-quark
Yukawa coupling $\ytop$ which might well be of order $O(1)$.
We postpone the discussion of
 its effects on our analysis to the end of this letter.

In the  approximation where all Yukawa couplings are set to zero,
the RG equations
 simplify as follows\foot{
For simplicity,  we also assume  $A_{ij}$ to be symmetric in flavor space and
neglect
$S=m^2_{h_u}-m^2_{h_d}+Tr(m^2_q-m^2_l-2m_u^2+m^2_d+m^2_e)$ whose coeffient
in the RG-eq. is small.
}
$$
\eqalign{
\dt \mg_a\ =&\  - \coeff{1}{4\pi} b_a \alpha_a \mg_a,
\qquad b_a = (11,1,-3)\ , \cr
\dt (m^2_q)_{ij}\ =& \ \coeff{\delta_{ij}}{4 \pi} (
\coeff{16}{3} \alpha_3 \mg^2_3  + 3 \alpha_2 \mg^2_2  + \coeff19 \alpha_1
\mg^2_1)
 -  \coeff{1}{16\pi^2}\,  \mgrav^2((\Au)^2_{ij} + (\Ad)^2_{ij}) \ , \cr
\dt (m^2_u)_{ij}\ =& \  \coeff{\delta_{ij}}{4 \pi}
 (\coeff{16}{3} \alpha_3 \mg^2_3   + \coeff{16}{9} \alpha_1
\mg^2_1)
 -  \coeff{1}{8\pi^2}\,  \mgrav^2 (\Au)^2_{ij} \ , \cr
\dt (m^2_d)_{ij}\ =& \ \coeff{\delta_{ij}}{4 \pi} (
\coeff{16}{3} \alpha_3 \mg^2_3    + \coeff49 \alpha_1
\mg^2_1)
 -  \coeff{1}{8\pi^2}\,  \mgrav^2(\Ad)^2_{ij} \ , \cr
\dt (m^2_l)_{ij}\ =& \ \coeff{\delta_{ij}}{4 \pi}
(3 \alpha_2 \mg^2_2  + \alpha_1 \mg^2_1)
 -   \coeff{1}{16\pi^2}\, \mgrav^2 (\Al)^2_{ij} \ , \cr
\dt (m^2_e)_{ij}\ =& \ \coeff{\delta_{ij}}{4 \pi} (
4\alpha_1  \mg^2_1)
 -  \coeff{1}{8\pi^2}\,   \mgrav^2(\Al)^2_{ij} \ , \cr
\dt  A^{u}_{ij}\ =&\ \coeff{1}{4\pi}
(\coeff83 \alpha_3 + \coeff32 \alpha_2 + \coeff{13}{18}\alpha_1)
A^{u}_{ij}\ , \cr
\dt  A^{d}_{ij}\ =&\ \coeff{1}{4\pi}
(\coeff83 \alpha_3 + \coeff32 \alpha_2 + \coeff{7}{18}\alpha_1)
A^{d}_{ij}\ , \cr
\dt  A^{l}_{ij}\ = &\ \coeff{1}{4\pi}
(\coeff32 \alpha_2 + \coeff{3}{2} \alpha_1) A^{l}_{ij} , \cr
}
\eqn\RGscalar
$$
where  $t= 2 \ln (\Mgut/ Q) $.
The solutions of
eqs.~\RGscalar\  determine the
 mass parameters at low energies ($Q = \Mweak$) in terms of their
boundary values at the high energy scale $\Mgut$.
   The important  point  in eqs.~\RGscalar\
is  the different renormalization of the diagonal and  off-diagonal
mass terms.
The diagonal  matrix elements  become larger at low energies
if the gauginos are sufficiently heavy   \refs{\DKS, \BIM}.
 On the other hand the
off-diagonal mass terms can
renormalize down if off-diagonal $A$-terms are present.
Thus their ratio  can be significantly smaller at low energies
than their boundary value at $\Mgut$.
This observation is the main physical effect we want to study.

Additional contributions to the low energy masses
arise from electroweak symmetry breaking and induce mass terms mixing
left and right handed scalars. Altogether, the physical masses
 of the squarks and sleptons at $\Mz$
appear in
three $3\times 3$ mass matrices
 $\Mm^{2\, (f)}_{LL}, \Mm^{2\, (f)}_{RR}, \Mm^{2\, (f)}_{LR}$
which together form the $6\times 6$  mass matrix
$$
M^{2 \ (f)}\   =\
\left( \matrix{  \Mm^{2 \ (f)}_{LL} & \Mm^{2 \ (f)}_{LR}\cr
                         \Mm^{2 \ (f)}_{LR}& \Mm^{2 \ (f)}_{RR} }\right) \ .
\eqn\smasses
$$
There is one such matrix
for each of the squarks
(up and   down)
as well as the sleptons
 (\ie\ $f  = u,d,l$).

We have already observed that
 the RG-equations \RGscalar\
are different
for the  diagonal and off-diagonal elements of the scalar mass matrix.
Let us therefore denote the off-diagonal  elements ($ i \neq j$) at $\Mweak$
by $\DMm^2_{ij}$  while $\Delta m^2_{ij}$ indicates the off-diagonal terms at
$\Mgut$.
The solutions of eqs.~\RGscalar\  determine $\DMm^2$
in  terms of  the high energy input parameters $\Delta m^2$ and  $A$
as follows
$$
\eqalign{
(\DMm^{2 \ (u)}_{LL})_{ij}  &= (\DMm^{2 \ (d)}_{LL})_{ij} =
\ (\Delta m^2_{q})_{ij}
-  \mgrav^2\,  \left( 1.8\ (\Au)^2_{ij}  + 1.7\ (\Ad)^2_{ij} \right) \ , \cr
(\DMm^{2 \ (u)}_{RR})_{ij}  &=
\ (\Delta m^2_{u})_{ij}
 -  3.6\ \mgrav^2\,  (\Au)^2_{ij}    \ , \cr
(\DMm^{2 \ (d)}_{RR})_{ij}  &=
\ (\Delta m^2_{d})_{ij}
-  3.4\ \mgrav^2\,  (\Ad)^2_{ij}   \ , \cr
(\DMm^{2 \ (l)}_{LL})_{ij} & =
\ ( \Delta m^2_{l})_{ij}
-  0.7\ \mgrav^2\,  (\Al)^2_{ij}   \ , \cr
(\DMm^{2 \ (l)}_{RR})_{ij} & =
\ (\Delta m^2_{e})_{ij}
-  1.4\ \mgrav^2\,  (\Al)^2_{ij}   \ , \cr
(\DMm^{2 \ (u)}_{LR})_{ij} & = \
3.7\ \mgrav\, \Au_{ij} \ \vev{\hu} \ ,\cr
(\DMm^{2 \  (d)}_{LR})_{ij} & = \
3.6\ \mgrav\, \Ad_{ij}\ \vev{\hd} \ ,\cr
(\DMm^{2 \ (l)}_{LR})_{ij} &=  \
1.5\ \mgrav\, \Al_{ij}\ \vev{\hd} \  .\cr
}
\eqn\msol
$$
(The numerical coefficients  have been obtained for
$\alpha_{\rm GUT}= 1/24$,
 $\Mgut = 3.6 \times 10^{16}\, GeV, Q = 100\, GeV $;
 $\vev\hd, \vev\hu$ denote the VEVs of the two Higgses and we
 use the typical values $\vev\hd = 60\, GeV, \vev\hu = 150\, GeV$
($\tan\beta = 2.5$) throughout this paper.)
We  see that at low energies $\DMm^2$   can be
very small (even tuned to zero) if  non-universal $A$-terms
of order $O(1)$ are present at $\Mgut$.

The renormalization
of the diagonal mass terms  is conveniently discussed by
defining  (low energy)
 `average squark masses'  $\mav^{2\ (f) }$
$$
\mav^{2\ (f) }\equiv
\coeff16  (\Tr \ M^{2 \, (f) }_{LL} + \Tr \ M^{2 \, (f) }_{RR} ) \ ,
\eqn\defmav
$$
and the corresponding average masses at high energies
$$
\eqalign{
\mavzero^{2\ (u)}  &\equiv \coeff16 (\Tr \ m^2_{q} \ + \Tr \ m^2_{u})
\ , \cr
\mavzero^{2\ (d)}  &\equiv \coeff16 (\Tr  \ m^2_{q} \ + \Tr \ m^2_{d})
\ , \cr
\mavzero^{2\ (l)}  &\equiv \coeff16 (\Tr \ m^2_{l} \ + \Tr \ m^2_{e})
\ . \cr
}
\eqn\devmavzero
$$
The solutions of eqs.~\RGscalar\ establish their connection
$$
\eqalign{
\mav^{2\ (u)}\ &=\  \mavzero^{2\ (u)} \, +\,  7 \, \mgzero^2\, -\,
 \mgrav^2  \left( 0.9\,\Tr\, (\Au)^2 + 0.3\, \Tr\, (\Ad)^2\right)
+ O(M_Z^2)
\ , \cr
\mav^{2\ (d)} \ &=\  \mavzero^{2\ (d)} \, +\,  7 \, \mgzero^2\, -\,
 \mgrav^2  \left( 0.3\,\Tr\, (\Au)^2 + 0.9\, \Tr\, (\Ad)^2\right)
+ O(M_Z^2)\ ,   \cr
\mav^{2\ (l)}\ &=\  \mavzero^{2\ (l)} \, +\,  0.3 \, \mgzero^2\, -\,
   0.3\,\mgrav^2\Tr\, (\Al)^2 + O(M_Z^2)  \ , \cr
}
\eqn\mavrel
$$
where the $O(\Mweak)$
contributions arise from electroweak symmetry breaking.
Note that the $A$-terms as well as
$\mgzero$ appear in eqs.~\msol\ and \mavrel\  due to  renormalization
effects.
Eqs.~\mavrel\  show the possible enhancement of $\mav^{(u,d)}$
at low energies
due to a large gaugino mass. However, this effect is
 much weaker in the slepton sector  since here
the renormalization is driven by $\alpha_{2}$ instead of $\alpha_3$
(see eqs.~\RGscalar).

Now we are prepared to discuss the constraints imposed by \bounds\ on
the low energy scalar mass matrices.
Apart from the SM contribution there exist additional
 contributions to FCNC processes induced
by  gaugino exchange in box and penguin diagrams.
We do not repeat  the explicit  computation here   but
instead  just summarize (and update) the relevant results
obtained in refs.~\refs{\GM, \HKT}.
The calculation can be performed in  two different
physically equivalent  sfermion bases.
For our purpose it is most convenient to  use a basis where the
fermion masses are diagonal,  the
gaugino-sfermion-fermion
couplings are diagonal
and the sfermion masses are arbitrary.\foot{
For non-zero Yukawa couplings this implies
 a CKM rotation on the scalar mass matrices
in order to keep the gaugino-sfermion-fermion
couplings  diagonal.}
The experimental bounds \bounds\  constrain
$\DMm^2$ and are conveniently expressed in terms of  \NS
$$
(\delta_{MN}^{(f)})_{ij}   \equiv
  {(\DMm^{2 \ (f)}_{MN})_{ij} \over \mav^{2\ (f)}} \ , \qquad
\vev{\delta^{(f)}_{ij}  }\equiv
 \sqrt{(\delta^{ (f)}_{LL})_{ij} (\delta^{(f)}_{ RR})_{ij}} \ ,
\eqn\Kresult
$$
where $M,N$ each  take the values $L,R$.
 In table~1  we display  the constraints on $\delta$ implied by
\bounds\  ({\it cf.} \refs{\GM, \NS}).
The numerical values are
obtained from eqs.~\bounds\ and the formulas
given in ref.~\GM.\foot{
We thank F.~Gabbiani for his assistance with some aspects of ref.~\GM.}
The relevant Feynman diagrams   depend on the gaugino masses (at $\Mweak$)
via the ratios  $x^{(u, d)}= \mg_3^2 (\Mweak)/ \mav^{2\, (u, d)}$
 in the squark sector and
$x^{(l)} = \mg_1^2 (\Mweak)/ \mav^{2\, (l)}$ in the slepton sector.
However, the dependence on $x^{(f)}$ is rather weak and in table~1
the typical values
  $x^{(u)}=x^{(d)}=1$,  $x^{(l)}=0.5$ have  been used.\foot{
We will see  that the values
chosen for $x^{(f)}$  are also phenomenologically sensible.}
In the squark sector
  large hadronic uncertainties  enter the estimates
and
 the values of  table~1  are only correct up to factors of $O(1)$.
The slepton sector is not plagued by such uncertainties; however,
here the relevant Feynman diagrams depend on the diagonal
$A$-terms as well as the supersymmetric $\mu$-parameter (see ref.~\GM).
For a reasonable range of parameters the constraints in the
slepton sector can vary by an order of magnitude.
In table~1 we have chosen an appropriate average value but
similar to the squark sector the numerical numbers should only be trusted
up to factors of $O(1)$.
Furthermore, only gaugino exchange diagrams
are taken into account whereas all
other contributions such as the SM diagrams
and their supersymmetric analogues  are neglected.
In most cases this is justified since they are subleading
and the leading gaugino contributions are only known within an
accuracy of $O(1)$.
However, in the up-squark sector for $\DMm_{13}^{2\, (u)}$ and
 $\DMm_{23}^{2\, (u)}$
the electroweak contributions become the leading constraint.
Evaluation of the relevant electroweak diagrams \Anja\
shows that they are suppressed by at least an order of magnitude
compared to the
constraints arising from $B-\bar B$ and $D-\bar D$ mixing
and therefore are neglected in the analysis below.
(A more detailed discussion of their possible effects can be
found in ref.~\Frank.)

%
%
{\bf 3}.
 It is instructive to split the  analysis into two separate cases.
First we concentrate on the situation where all $A$-terms are small
and (together with the Yukawa couplings) can be neglected in the
RG-analysis.\foot{
This essentially repeats and extends the analysis of ref.~\DKS.}
{}From eqs.~\msol\ we  learn that for vanishing $A$-terms all
 constraints on
$\DMm^{2\, (f)} _{LR}$ are  automatically satisfied.
Furthermore, the $\DMm^2$'s  do not renormalize
and are directly
determined by their boundary values at
$\Mgut$ (see eqs.~\msol).
Thus,
 each entry of table~1   can be translated
into a constraint
on a three-dimensional soft parameter space (at $\Mgut$)
spanned by
$(\mavzero, \mgzero, \Delta m^2)$.
Let us assume    $\Delta m^2 \simeq \mavzero^2$
(all matrix elements are of the same order of magnitude) and
 study the strongest constraints  in the squark and
slepton sector which
 arise from $K-\bar K$ mixing and
 $\mu\to e \gamma$.
{}From eqs.~\Kresult, \msol\ and \mavrel\ we learn
$$
\eqalign{
(\delta^{(d)}_{RR})_{12} = (\delta^{(d)}_{LL})_{12}
&={ 1  \over 1 + 7\ \xzero^{(d)}} \ , \qquad
  (\delta^{(l)}_{RR})_{12} = (\delta^{(l)}_{LL})_{12}
={1 \over 1 + 0.3\ \xzero^{(l)}}\, , \cr
&{\rm where} \quad \xzero^{(f)}
\equiv\ {\mgzero^2 \over \mavzero^{2\, (f)} }\, .
}
\eqn\ddc
$$
 In fig.~1 we display the constraints on the ratio
$\sqrt\xzero$ as a function
of the (physical) low energy average squark (slepton)
mass $\mav$.\foot{
Both constraints are shown in the same plot for better comparison,
strictly speaking there is a different $(\mav^{(d)}, \xzero^{(d)})$ and
 $(\mav^{(l)}, \xzero^{(l)})$
in each case.}
We see that the constraints can be satisfied if the gaugino mass $\mgzero$
is significantly larger than $\mavzero$.
By far the strongest constraint arises in the slepton sector
from $\mu\to e \gamma$
and for small  $\mav$ a rather large hierarchy
is required ($\sqrt\xzero \gsim 30)$.
For the squarks this ratio can be  lower ($\sqrt\xzero \gsim 10$ )
 due to the large renormalization effect
induced by  $\mgzero$ (eq.~\mavrel).\foot{
Note that with such a big hierarchy
 $\mav$  is almost entirely determined by $\mgzero$ and
$\mavzero  \simeq 0$ for small $\mav$.}
For higher values of $\mav$  the slepton constraint
falls off faster than  the squark constraint
due to the different scaling behavior of
the penguin versus the box diagram  (see table~1) and for large
$\mav$ ($1 \ TeV$) one   needs $\sqrt\xzero \gsim 6$
in both sectors.
However, we should
stress  that fig.~1 should only be trusted up to factors of $O(1)$
due to the uncertainties implicit in table~1.

The ratio of the gaugino masses to the
squark (slepton)  masses at low energies
does not require any  hierarchy.
The gaugino masses renormalize according to
$\mg_3 (\Mweak)\ \simeq\ 3\,  \mgzero, \,
\mg_1 (\Mweak)\ \simeq\ 0.4\,  \mgzero$
which implies
$$
x^{(d)} \equiv {\mg_3^2 (\Mweak) \over \mav^{2\, (d)}}
\simeq  {9\, \xzero^{(d)} \over 1+7\, \xzero^{(d)}} \
\to
{9\over 7}  ,
\quad
x^{(l)}  \equiv  {\mg_1^2 (\Mweak)\over \mav^{2\, (l)}}
\simeq   { 0.16\, \xzero^{(l)} \over 1+ 0.3\, \xzero^{(l)}} \to 0.53 .
\eqn\xrel
$$
Thus $x^{(d)}$ only
 takes values  in the interval $[0,\coeff97)$ and for
 large $\mgzero$ (large $\xzero$) approaches $\coeff97$.
Similarly, $x^{(l)} $ takes values in  $[0,0.53)$
and hence the photino is  always lighter than the sleptons at $\Mweak$.
The fact that the ratios $x^{(f)} $ at low energies approach
a fixed point for large $\xzero$
is another reason for choosing $x^{(d)}=1, x^{(l)}= \coeff12$ in  table~1.

So far we have observed that in order to satisfy  the constraints
\bounds\ a large hierarchy (at $\Mgut$) between $\mgzero$ and $\mavzero$
is required.
However, this has been obtained under the assumption that
all matrix elements of  the scalar mass matrix are of the same order of
magnitude.
{}From eqs.~\ddc\ and table ~1 it follows
that an   `inverse' hierarchy $\xzero \ll 1$   requires
$$
{\Delta m^2_{12} \over \mavzero^{2\, (d)} }
\simeq 2\cdot 10^{-2}  \left({\mav^{(d)} \over 1 TeV}\right) \ , \qquad
{\Delta m^2_{12} \over \mavzero^{2\, (l)} } \simeq 0.3
\left({\mav^{(l)}\over 1 TeV}\right)^2 \ ,
\eqn\inversh
$$
\ie\ some degree of universality of the scalar masses.

%
{\bf 4}.
Let us  turn to the case where also non-universal $A$-terms are
 kept in the RG-evolution. {}From eqs.~\msol\ we learn
 that the constraints on $\DMm^{2\, (f)} _{LR}$
directly translate into constraints on the off-diagonal
$A$-terms.
Table~1 implies that   $K-\bar K$ mixing and
 $\mu\to e \gamma$  constrain
   $\Ad_{12}$ and $\Al_{12}$ severely whereas
the constraint on $\Au_{12}$ is significantly weaker and
the diagonal terms $A^{(f)}_{11}, A^{(f)}_{22}$ are
unconstrained by the bounds \bounds.
To simplify the analysis
we impose
$A^{(f)}_{12} \simeq A^{(f)}_{11} \simeq A^{(f)}_{22} \simeq 0$
and choose the remaining  $A$-terms in each sector to be
 identical, \ie\  the $A$-matrices
 have the `texture'\foot{
This texture is also suggested in some superstring models
\refs{\KLa,\LN}
where typically $A^f \simeq Y^f_{33}$ holds.
This implies  large quantum corrections
to the fermion masses \Banks\ and  opens up the possibility
of generating the entire fermion mass hierarchy from $A$-terms
as a quantum effect.
Work along these lines is in progress.
}
$$
A^{(f)}\   =\  \left( \matrix{  0     & 0       & A^{f}\cr
                                    0      & 0       & A^{f}\cr
                                   A^{f}  & A^{f}   & A^{f}}
\right) \ .
\eqn\Atexture
$$
As before, we take the off-diagonal elements of the
scalar mass matrices to be of the same order of magnitude
$\Delta m^2 = \mavzero^{2} = \mgrav^2$ and hence the
 soft parameter space
is spanned by
$(\mavzero, \xzero, \Au, \Ad)$ in the squark sector
and
$(\mavzero, \xzero, \Al)$ in the slepton sector.

{}From eqs.~\msol\ we observe that $A$-terms do renormalize
the off-diagonal scalar mass terms such that
they become smaller at low energies. Thus we expect the the bounds \bounds\
to be more easily satisfied when non-universal $A$-terms are present.
Using eqs.~\msol, \mavrel, \Atexture\
(and neglecting terms of $O(M_Z^2)$
as before) we find in the slepton sector
$$
\eqalign{
(\delta^{(l)}_{LL})_{12} = \, &
(\delta^{(l)}_{LL})_{13} =
(\delta^{(l)}_{LL})_{23} = \
{1 - 0.7\, (A^l)^2 \over 1 - 1.5\,  (A^l)^2 + 0.3\, \xzero^{(l)} }\, ,
\cr
(\delta^{(l)}_{RR})_{12} = \,  &
(\delta^{(l)}_{RR})_{13} =
(\delta^{(l)}_{RR})_{23} = \
{1 - 1.4\, (A^l)^2 \over 1 - 1.5\,  (A^l)^2 + 0.3\, \xzero^{(l)} }\, ,
\cr
(\delta^{(l)}_{LR})_{12} =\,  & 0\ , \quad
(\delta^{(l)}_{LR})_{13} = (\delta^{(l)}_{LR})_{23} =\
 {1.5\,  A^l\,  \vev\hd \over
\mavzero (1 - 1.5\,  (A^l)^2 + 0.3\, \xzero^{(l)}) }\ .
}
\eqn\cons
$$
We see that  for
 $A^l$ of $O(1)$ one of  the numerators in $\delta_{LL}$ or $\delta_{RR}$
can be made small and even tuned to zero\foot{
A similar situation occurs in  the supersymmetric Coleman-Weinberg
mechanism where a Higgs parameter is tuned negative \COWE.}
but not both simultaneously.
In addition, one has to make sure that $\mav^{(l)}> 50\, GeV$
or in other words the denominators in eqs.~\cons\ stay positive.
This again requires a hierarchy between gaugino and slepton masses
 $\xzero^{(l)} > 1$.
The absolute value of the numerators in $\delta_{LL}$ and $\delta_{RR}$
should  both be small and
for fixed  $\xzero^{(l)}$  this  leads to a lower and  upper bound
for $A^l$.
Furthermore, the constraints for $\delta_{LR}$   and $\mav^{(l)}> 50\, GeV$
also imply   upper  bounds on $A^l$.
Together, the upper and lower bounds lead to a `wedge' shaped
curve which we show   in fig.~2
as a function of the low energy
slepton mass $\mav^{(l)}$ for fixed $\sqrt{\xzero^{(l)}} = 8, 16$.
Leaving   $\xzero^{(l)}$ arbitrary one can scan over the parameter space
$(\mavzero^{(l)}, \xzero^{(l)}, \Al)$ and demand all constraints
in table~1 to be satisfied.
In fig.~3 we display the lowest allowed $\xzero^{(l)}$ as a function of
$\mav^{(l)}$   for an arbitrary $A^l$  and for fixed
 $A^l = Y_\tau$ (which essentially coincides with
 $A^l=0$).
We indeed see that it is
 easier to obey the constraints for arbitrary $A^l$ and as a consequence
a smaller value for $\sqrt\xzero$ is tolerable.

In the squark sector  we perform a similar analysis which becomes somewhat
more
involved since we have two independent $A$-terms ($A^u, A^d$) to play with.
Using eqs.~\msol, \mavrel, \Atexture\
 we have
$$
\eqalign{
(\delta^{(d)}_{LL})_{12} =&\
(\delta^{(d)}_{LL})_{13} =
(\delta^{(d)}_{LL})_{23} =
{1 - (1.8\, (A^u)^2 + 1.7\, (A^d)^2  )
\over 1 - 1.5\,  (A^u)^2 - 4.5\, (A^d)^2 + 7\, \xzero }\, ,
\cr
(\delta^{(d)}_{RR})_{12} =&\
(\delta^{(d)}_{RR})_{13} =
(\delta^{(d)}_{RR})_{23} =
{1 - 3.4\,  (A^d)^2
\over 1 - 1.5\,  (A^u)^2 - 4.5\, (A^d)^2 + 7\, \xzero }\, ,
\cr
(\delta^{(d)}_{LR})_{12} =& 0\ , \quad
(\delta^{(d)}_{LR})_{13} =
(\delta^{(d)}_{LR})_{23} =
 {3.6\,  A^d\,  \vev\hd
\over \mavzero (1 - 1.5\,  (A^u)^2 - 4.5\, (A^d)^2 + 7\, \xzero )}\ ,
}
\eqn\consdown
$$
and a similar set in the up-squark sector.
We numerically scan over the parameter space ($A^u, A^d, \xzero, \mgrav$) and
check that all constraints in table~1 are satisfied simultaneously.\foot{
We also check  that the lowest eigenvalue of the squark mass matrix is above
$100\, GeV$.}
In fig.~4 we  display
the lowest possible $\sqrt\xzero$ as a function  of the physical
(low energy) squark mass $M^d_{av}$
for
arbitrary (allowed) $A^u, A^d$,
for  fixed $A^u = \ytop,\,  A^d = Y_b$ and also for  $A^u= A^d =0$.
As expected for non-zero $A$
the allowed values for $\sqrt\xzero$ are significantly
lower than for $A=0$  but we again need
 $\sqrt\xzero >1$.
For fixed $\xzero$ the $A^u$ and $A^d$ satisfy again upper and lower
bounds in analogy with  the slepton sector.

Finally, the `inverse' hierarchy $\xzero < 1$  does not allow
large $A$-terms since $\mav$ would become too small.
Thus in this case one is forced  to the case of vanishing
$A$-terms and the situation
discussed in eq.~\inversh.

{\bf 5}.
So far the discussion has been in the limit of vanishing Yukawa couplings.
 Turning on the Yukawa couplings
we first need to transform to the fermion mass basis with an appropriate
CKM rotation.
Since the constraints in table~1 are given in a basis where also
the fermion-sfermion-gaugino couplings are diagonal one needs to perform
 a `compensating' CKM rotation on the scalar degrees of freedom.
It is possible that after such a rotation the scalar mass matrices are
automatically diagonal.
 This `squark-quark alignment mechanism'
has been recently discussed in ref.~\NS\ and shown to arise naturally in
theories
with horizontal symmetries.
Here we assume that no such alignment occurs and the CKM-rotated
mass matrix is still arbitrary with $\Delta m^2 = O(\mgrav^2)$.
In the RG-analysis we now keep the top-quark Yukawa coupling
$\ytop$ which is the leading effect.\foot{
For large $\tan\beta$ one should keep
all Yukawa couplings of the third generation but we do not discuss
this case here.}
It is important to note that in this
 approximation
the  fermion mass basis  does not rotate between
 $\Mweak$   and $\Mgut$ and one
stays in the fermion mass basis chosen at $\Mgut$ along the RG-trajectory.

For universal $A$-terms a non-zero $\ytop$ only affects the renormalization
of $\mav^{(u,d)}$ and only via the renormalization of $(m^2_q)_{33}$
and $(m^2_u)_{33}$.
This changes the coefficient of $(m^2_q)_{33}$ and  $(m^2_u)_{33}$
in eqs.~\devmavzero, \mavrel\ by at most  a factor of 2 which gets weakened
by taking the average mass.
(Also the coefficient of $\mgzero$ changes slightly.)
Hence, for universal $A$-terms
 the effect of a large $\ytop$ is well within the (hadronic)
uncertainties of the constraints given in table~1
and hence fig.~1
 still holds in our approximation.

For non-universal $A$-terms the effect of a large
$\ytop$ is more significant
(the  RG-analysis can be found in ref.~\Frank).
Again  only  the squark sector is affected whereas the sleptons
(and hence figs.~2,3)
are unchanged.
The most important effect is that all coefficients in front of the
$A^u$-terms in eqs.~\consdown\ become $\ytop$ dependent  and for a large
$\ytop$ decrease significantly
(they approach 0 for $\ytop$ at its IR (quasi-) fixed point.)
Hence the `help' from the $A$-terms in satisfying the FCNC constraints
becomes weaker for large $\ytop$.
 In fig.~5 we display this effect for three different values
of $\ytop$
(for $\tan\beta = 2.5$ they correspond to
$m_{\rm top} = 0, 160, 175, 180\, GeV$).
When $\ytop$ approaches the fixed point
the constraint on $\sqrt\xzero$ moves towards the $A=0$ case.
However, as before the most stringent constraint arises
in the slepton sector
which is independent of $\ytop$.

{\bf 6}.
Let us summarize the main points of our analysis.
The smallness of FCNC imposes severe constraints on
the off-diagonal elements of the scalar mass matrices at low energies.
In translating these constraints to some high energy scale $\Mgut$
(which is of interest for model building) quantum corrections have
to be included and can (significantly)   change  the conditions
on the supersymmetric parameter space at $\Mgut$.
In particular, a  (large)   hierarchy between the (high energy)
 gaugino mass and the (high energy) scalar masses  ensures that
at low energies the FCNC constraints are satisfied even for
non-universal scalar masses at $\Mgut$.
Non-universal $A$-terms also dilute the constraints
at $\Mgut$ but one still needs the
gaugino mass to be larger than the  scalar masses.
An inverse hierarchy  with scalar masses  bigger than the gaugino mass
(again at $\Mgut$) is only possible if there is a (significant)
degree of universality within the scalar mass matrices.
By far the strongest constraints on the high energy parameter space arise
in the slepton sector of the theory.

\ack

We would like to thank Y.~Nir for many discussions and useful comments on
the manuscript.
 F.~Gabbiani and A.~Masiero  helped clarifying some aspects of ref.~\GM.
We are grateful to the Lehrstuhl Wagner for their assistance
in producing the figures.
The work of J.L.~is supported by a Heisenberg fellowship of the DFG.

\refout

\font\twelveib=cmmib10 scaled \magstep 1
\def\bmi{\textfont1=\twelveib \fam1 }

\def\mav{M_{\rm av}}
\bigskip
\bigskip
$$\offinterlineskip \tabskip=0pt
  \vbox{
  \halign {\vrule width 0pt height 14pt depth 4pt  \vrule width 1pt #     &
  \quad  \hfil # \quad &
  \vrule #     &
  \quad \hfil  # \quad &
  \vrule#     &
  \quad  \hfil # \quad &
  \vrule width 1pt #     \cr
  \noalign{\hrule height 1pt}
  & \multispan 5 \hfil $\bmi K\bar K$ \hfil &  \cr
  \noalign{\hrule}
  & $(\delta_{LL/RR}^{(d)})_{12}(x^{(d)}=1)$
&& $(\delta_{LR}^{(d)})_{12}(x^{(d)}=1)$
&& $\vev{\delta^{(d)}_{12}}(x^{(d)}=1)$ & \cr
  \noalign{\hrule}
  & $1\cdot 10^{-1} \mav /1TeV$ && $1\cdot 10^{-2} \mav /1TeV$
&&  $8\cdot 10^{-3} \mav /1TeV$ & \cr
  \noalign{\hrule height 0.7pt}
  & \multispan 5 \hfil $\bmi B\bar B$ \hfil & \cr
  \noalign{\hrule}
  & $(\delta_{LL/RR}^{(d)})_{13}(x^{(d)}=1)$
&& $(\delta_{LR}^{(d)})_{13}(x^{(d)}=1)$
&& $\vev{\delta^{(d)}_{13}}(x^{(d)}=1)$ & \cr
  \noalign{\hrule}
  & $2\cdot 10^{-1} \mav /1TeV$ && $5\cdot 10^{-2} \mav /1TeV$
&&  $3\cdot 10^{-2} \mav /1TeV$ & \cr
  \noalign{\hrule height 0.7pt}
  & \multispan 5 \hfil $\bmi D\bar D$ \hfil & \cr
  \noalign{\hrule}
  &  $(\delta_{LL/RR}^{(u)})_{12}(x^{(u)}=1)$
&& $(\delta_{LR}^{(u)})_{12}(x^{(u)}=1)$
&& $\vev{\delta^{(u)}_{12}}(x^{(u)}=1)$  &\cr
  \noalign{\hrule}
  & $2\cdot 10^{-1} \mav /1TeV$ && $5\cdot 10^{-2} \mav /1TeV$
&&  $3\cdot 10^{-2} \mav /1TeV$ & \cr
  \noalign{\hrule height 0.7pt}
  & \multispan 5 \hfil $\bmi b\rightarrow s\gamma $ \hfil &\cr
  \noalign{\hrule}
  & $(\delta_{LL/RR}^{(d)})_{23}(x^{(d)}=1)$
&& $(\delta_{LR}^{(d)})_{23}(x^{(d)}=1)$ && \hfil $-$ \hfil & \cr
  \noalign{\hrule}
  & $3\cdot 10^2 \  \mav ^2/(1TeV)^2$
&& $4\cdot 10^{-1} \mav /1TeV$ &&  \hfil $- $ \hfil & \cr
  \noalign{\hrule height 0.7pt}
  & \multispan 5 \hfil $\bmi \mu\rightarrow e\gamma$ \hfil &\cr
  \noalign{\hrule}
  & $(\delta_{LL/RR}^{(l)})_{12}(x^{(l)}=0.5)$
&& $(\delta_{LR}^{(l)})_{12}(x^{(l)}=0.5)$ && \hfil $-$ \hfil &\cr
  \noalign{\hrule}
  & $ 1\cdot 10^{-1} \ \mav ^2/(1TeV)^2$
&& $2\cdot 10^{-5} \mav /1TeV$ &&  \hfil $- $ \hfil & \cr
  \noalign{\hrule height 0.7pt}
  & \multispan 5 \hfil $\bmi \tau\rightarrow e\gamma$ \hfil &\cr
  \noalign{\hrule}
  & $(\delta_{LL/RR}^{(l)})_{13}(x^{(l)}=0.5)$
&& $(\delta_{LR}^{(l)})_{13}(x^{(l)}=0.5)$ && \hfil $-$ \hfil &\cr
  \noalign{\hrule}
  & $4\cdot 10^{3} \mav ^2/(1TeV)^2$ && $2 \ \mav /1TeV$
&&  \hfil $- $ \hfil & \cr
  \noalign{\hrule height 0.7pt}
  & \multispan 5 \hfil $\bmi \tau\rightarrow \mu\gamma$ \hfil &\cr
  \noalign{\hrule}
  & $(\delta_{LL/RR}^{(l)})_{23}(x^{(l)}=0.5)$
&& $(\delta_{LR}^{(l)})_{23}(x^{(l)}=0.5)$ && \hfil $-$ \hfil &\cr
  \noalign{\hrule}
  & $7\cdot 10^{2} \mav ^2/(1TeV)^2$
&& $2\cdot 10^{-1}  \ \mav /1TeV$ &&  \hfil $- $ \hfil & \cr
  \noalign{\hrule height 1pt}
  }}$$

 \centerline{ Table 1 :
{\it constraints from FCNC processes} \ \ \ \ \ \ \ \ \ \ \ \ \ }

\vfill\break

\noindent {\bf Figure Captions}

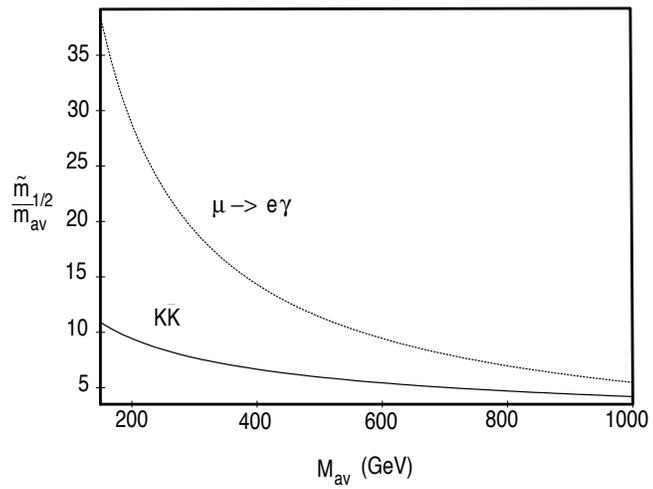
\noindent Figure 1: Constraints on the ratio
${\sqrt\xzero}= {\mgzero\over \mavzero} $ as a function
of the (low energy) average
mass $\mav^{(l)}$.

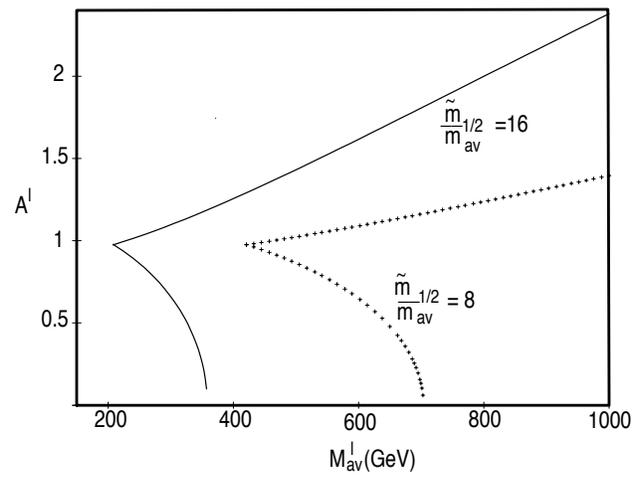
\noindent Figure 2: Upper and
lower bounds
on $A^l$  as a function of the
 average mass $\mav^{(l)}$ for a fixed
 $\sqrt\xzero = 8, 16$.

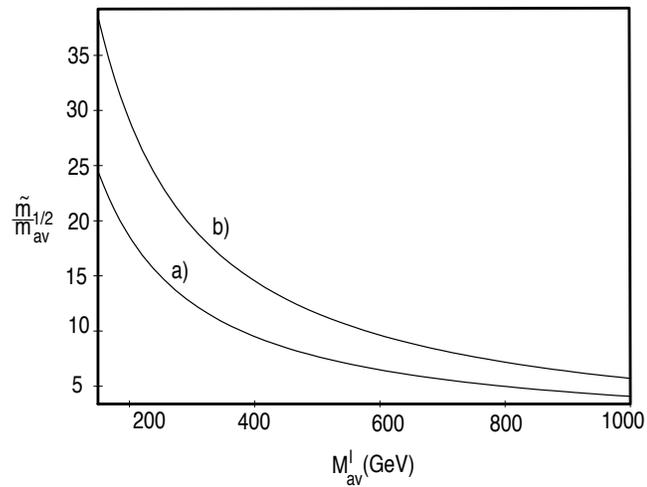
\noindent Figure 3: The lowest allowed $\sqrt\xzero$ as a function of
$\mav^{(l)}$ \ for  a) arbitrary $A^l$;
b) $A^l = Y_\tau$.

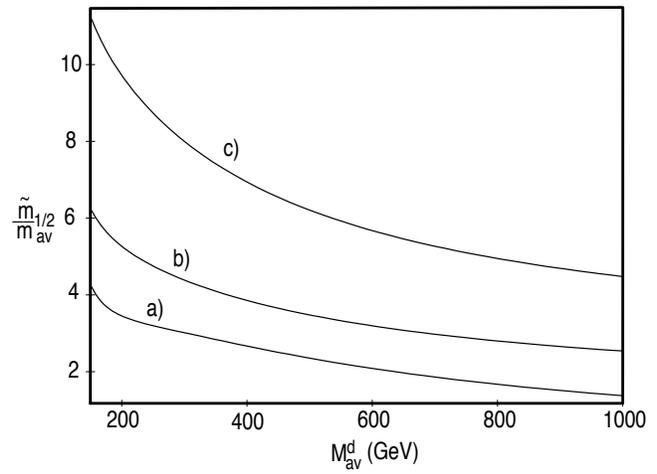
\noindent Figure 4: The minimal allowed $\sqrt\xzero$ as a function  of the
 average squark mass $M^{(d)}_{av}$
for  a)
arbitrary  $A^u, A^d$;
b) $A^u = \ytop, A^d = Y_b$;  c)  $A^u= A^d =0$.

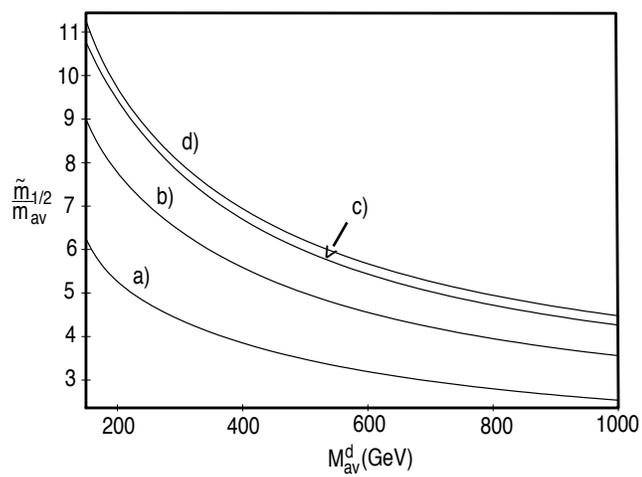
\noindent Figure 5: The minimal allowed $\sqrt\xzero$ as a function  of
$M^{(d)}_{av}$  for $\tan\beta=2.5$ and
 a) $m_{\rm top} = 0$;
 b) $m_{\rm top} = 160\, GeV$;
 c) $m_{\rm top} = 175\, GeV$;
 d) $m_{\rm top} = 180\, GeV$.
Case d) coincides with $A=0$.

\end